\documentstyle[11pt,aasms4]{article}
\begin{document}
\baselineskip 18.0pt
\def\oneskip{\vskip\baselineskip}
\def\xr#1{\parindent=0.0cm\hangindent=1cm\hangafter=1\indent#1\par}
\def\la{\raise.5ex\hbox{$<$}\kern-.8em\lower 1mm\hbox{$\sim$}}
\def\ma{\raise.5ex\hbox{$>$}\kern-.8em\lower 1mm\hbox{$\sim$}}
\def\ea{\it et al. \rm}
\def\am{$^{\prime}$\ }
\def\as{$^{\prime\prime}$\ }
\def\msol{M$_{\odot}$ }
\def\kms{$\rm km\, s^{-1}$}
\def\cm3{$\rm cm^{-3}$}
\def\Ts{$\rm T_{*}$}
\def\Vs{$\rm V_{s}$}
\def\n0{$\rm n_{0}$}
\def\B0{$\rm B_{0}$}
\def\ne{$\rm n_{e}$}
\def\Te{$\rm T_{e}$}
\def\Ne{$\rm N_{e}$}
\def\Tgr{$\rm T_{gr}$}
\def\Tgas{$\rm T_{gas}$}
\def\Ec{$\rm E_{c}$}
\def\erg{$\rm erg\, cm^{-2}\, s^{-1}$}
\def\Hb{H$\beta$}
\def\upa{$\uparrow$}
\def\dop{$\downarrow$}

\centerline{\Large{\bf Modeling RR Tel through the Evolution of the Spectra}}

\bigskip

\bigskip

\bigskip

\centerline{Marcella $\rm Contini^1$ and Liliana $\rm Formiggini^{2}$}

\bigskip

\bigskip

$^1$ School of Physics and Astronomy, Tel-Aviv University, Ramat-Aviv, Tel-Aviv,
69978, Israel

\bigskip

$^2$ Wise Observatory, The Raymond and Beverly Sackler Faculty of Exact Sciences,
Tel-Aviv University, Tel-Aviv, 69978, Israel

\bigskip

\bigskip

\bigskip

\bigskip

\bigskip

Running title : The evolution of RR Tel

\bigskip

\bigskip

\bigskip

Subject headings: binaries:symbiotic - shock waves - 

stars: individual: (RR Telescopii)

\newpage

\section*{Abstract}

We investigate the evolution of RR Tel after the outburst
by fitting the emission spectra in two epochs. The first one (1978) is characterized
by large fluctuations in the light curve and the second one (1993) by  the
slow fading trend. 
In the frame of a  colliding wind model two shocks are present:
the reverse shock  propagates in the direction of the white dwarf  and the
other one expands towards or beyond the giant.
The results of our modeling show that in 1993 the expanding shock has
overcome the  system and is  propagating in the nearby ISM.
The  large fluctuations observed in the 1978 light curve result
from  line intensity  rather than from  continuum variation.
These  variations are explained by fragmentation of matter at the time of
head-on collision of the winds from the two stars.
A high velocity (500 \kms) wind component is revealed from the fit
of the SED of the continuum in the X-ray range in 1978,
but is quite  unobservable in the line profiles.
The geometrical thickness of the emitting clumps is
the critical parameter which can explain the short time scale
variabilities of the spectrum and the trend of slow line
intensity decrease.

\newpage

\section{Introduction}

RR Tel belongs to the group of symbiotic novae. They  are distinguished
from  classic symbiotic stars by having undergone only one single
 major outburst and have not yet recovered. 
The outburst of  RR Tel started in 1944 when the system brightened by 7 mag.
 The light curve after the outburst shows a  gradual fading which is still
in progress. Besides the general decrease, the light curve is marked
by  periods of large fluctations.
The system is believed to be a binary, consisting of a nuclear
burning white dwarf (WD) and a late type companion, and  an emission
nebula  photoionized  by the hot star. The cool component  is a
Mira-type variable, with a pulsation period
of 387 days (Feast et al. 1983). Weak TiO absorption bands of spectral
type about M5 were observed by Thackeray (1977). 
The system has been  monitored in many different spectral regions,
ranging from radio to X-rays, and, recently, was
observed by the HST.
In spite of being extensively studied (Aller et al 1973, Thackeray 1977,
Hayes \& Nussbaumer 1986, etc.), many  questions about RR Tel remain 
unsolved.

The emission line spectra of symbiotic stars are usually calculated using a 
classical model where the main excitation mechanism is photoionization
from the hot star (Murset  et al 1991). However, in the last few years, the
importance of the  winds from both components in symbiotic
systems and their collision  has been addressed by many authors
 (e.g.  Wallerstein et al 1984,
 Girard \& Willson 1987, Nussbaumer \& Walder 1993, etc).

Recently a  model has beeen  successfully applied to the 
calculation of the line  and continuum spectra of HM Sge 
(Formiggini, Contini, \& Leibowitz 1995) and AG Peg (Contini 1997).
This model is able to fit the high and low ionization emission lines
by taking into account both the shock created by the winds and the
photoinization flux from the hot star.

In this paper we investigate the  evolution of RR Tel through the
interpretation of both line and continuum spectra 
in different epochs. Both the photoionizing flux from the hot star 
and the collision of the winds are considered, therefore the calculation
was performed with the  SUMA code.
SUMA (Contini  1997 and references therein) consistently 
accounts for both the radiation from the hot star and for the shock in 
a plane parallel geometry. 
Dust reradiation is also consistently calculated.

In \S 2 the evolution of the outburst is described. In \S 3 the
general model is presented. The spectra in 1978 and 1993 are
discussed in \S 4 and \S 5, respectively. 
Model results in the two epochs 1978 and 1993 are compared
in \S 6 and conclusions follow in \S 7.

\section{The evolution of the outburst}

\bigskip

Fig. 1 shows the visual light curve of RR Tel between the years 
1949 and 1995 which was obtained collecting data from various sources 
(AAVSO circulars, Heck \& Manfroid 1985, and Bateson 1995). 
A period of 387 days was observed (Heck \& Manfroid 1982) for
the system after 1930, the same which is observable in the infrared
(Feast et al 1983). It can be noticed that a general decreasing
trend started in 1949 and is still in progress. 
Above the fading trend the curve displays a decrease slightly larger 
than 0.5 mag in the years  1962 - 1963 and particularly large fluctuations 
in the years 1975-1982. During these years the cool
companion remained quite stable as shown by many observations collected at
infrared wavelengths where the cool star dominates.
In fact a periodicity of 387 days has been detected on JHKL photometric
data during 1975-1981 by Feast  et al (1983). After these events the system
recovered  the general fading trend.

Penston et al (1983)  reported optical photometric variations from night
to night and within a single night between 1975 and 1982. 
 We recall that the line spectrum is very rich, and that  some 
strong lines fall in the passbands of the broad band photometry.
It is then likely that large fluctuations in the line intensities  are
the source of the  strong variations in the visual light curve.
However, the question of whether the variation in the continuum or 
in the line intensities  are the cause of the photometric fluctuations,
is still  open. 

The outburst event led initially to an extended atmosphere
around the white dwarf. The nebular phase  emerged some years later
and is characterized by a wind with a terminal velocity
increasing from 400 \kms ~in 1949 to 1300 \kms ~in 1960 (Thackeray  1977).
However, when IUE became operational in 1978 Penston et al (1983)
measured a terminal velocity of about 75 \kms, indicating that the wind had
decelerated from initial velocities of thousands of \kms ~to less than 100 \kms.

As the outburst evolved, emission lines  from high ioniziation levels appeared
(Viotti 1988). The emission line spectra of RR Tel were  studied
by Penston et al (1983), Hayes \& Nussbaumer (1986), Aufdenberg (1993), etc.
Recently Zuccolo, Selvelli, \& Hack(1997) (from hereafter ZSH) retrieved
all the available IUE spectra up to 1993 and after an accurate
reduction and  correction for spectral artifacts, saturation  and other
instrumental effects, a homogenous and complete list of the emission 
lines in the UV range was  presented.

ZSH claim that the variations observed in the emission line 
intensities from 1978  to 1993 show a rather strong correlation
with the line ionization level. 
The decrease is  by a factor larger than 3 and up to 10 for low ionization
lines and between 2  and 3 for medium and high ionization lines. 
The fluxes for some representative lines from ZSH Table 1,  
are plotted in  Fig. 2. It can be noticed that  among the high
ionization level lines, only [MgVI] 1806 has increased between 1983 and 1993.
It seems that the conditions of the system did not change significantly.
Hayes \& Nussbaumer (1986) already noticed a lack of periodic
variation in the UV line fluxes over the period 1978-1984, finding  only
a slight decrease of the intensities.
A modulation of the line intensities with orbital phase is
observed in symbiotic stars (e.g. Kenyon et al 1993 for AG Peg).   
The lack of periodicity of the RR Tel line intensities over a 15
years span  (1978-1993) can be explained  either by the system being face-on
or by a period longer than 15 y. Orbital periods of symbiotic
Miras are  believed to be longer than 20 years (Withelock 1988).

Actually, for the  modeling purpose, we consider the system in two 
representative epochs, one in 1978  when the system underwent  high 
variability in the visual light on small time scale, 
and the other in 1993 after recovering  the slow fading trend.
We take advantage of the ZSH  collection of data for the  
fitting of model calculations.

\section{The general model}

A description of the colliding wind model is given by Girard \& 
Willson (1987).
Recall that the cool component of RR Tel is a Mira, characterized
by a strong mass loss rate that produces an extended atmosphere.
The outburst is the starting point of the wind from the hot star.
The winds from the two components approach from opposite directions
and collide head-on in the region between the stars and head-on-tail
outside.  It is plausible that fragmentation of matter in geometrically
thin clumps results from turbulence caused by  collision.
The WD ejecta propagate in the giant atmosphere towards higher
densities.  Due to the  collision  
two main shocks appear, the reverse shock and the expanding one. 
The reverse shock propagates back in the WD ejecta 
facing the WD. Therefore, the photoionizing radiation flux
from the WD reaches the very shock front. On the other hand,
the secondary shock expanding towards the red giant, propagates in a denser
medium, therefore, its velocity is lower than that of the reverse one.
In this case the radiation from the WD reaches the matter downstream,
opposite to the shock front. A simple sketch of the model in different
epochs is given in Figs. 3a and 3b. 

The initially high velocity of the WD wind will produce high densities and temperatures
behind the shock front, and hence hard X-ray emission would be
expected. However, after propagating and  sweeping up additional material
the shock decelerates. Lower shock velocities correspond to lower
temperatures downstream and the ionized region behind the
shock front will become visible in the UV and optical spectra.
In  our  model  of RR Tel two gas regions contribute to the emission spectra :
downstream of the reverse shock and downstream of the expanding one.

In this paper we consider the system in the two epochs  
1978 and  in 1993. The input parameters for the  calculation of the emitted
spectra by SUMA are the following : the shock velocity, \Vs; the
preshock density, \n0; the magnetic field, \B0; the colour temperature 
of the hot star, \Ts; the ionization parameter, U; the relative abundances
of the elements; and  d/g, the dust-to-gas ratio by number.
The  models are matter bound and the geometrical
thickness of the emitting clumps, d, is defined by the best results.

The choice of the input parameters is based on the observational 
evidence at the epochs selected for modeling. 
The densities are constrained by the critical line ratios 
(Nussbaumer \& Schild, 1981)
and the shock velocities by the FWHM of the line profiles.
For all models  \Ts = 1.4 $\times 10^5$ K 
(Murset et al. 1991)  and  \B0 = $10^{-3}$ gauss,  
which  is suitable  to red giant athmospheres (Bohigas et al 1989), are adopted.
 Relative abundances are calculated
phenomenologically within the range of values which are consistent
with CV systems.

\section{The spectra in 1978}

The  models which best fit the observation data are selected from a grid 
of model calculations.  Notice, however, that, 
whenever different regions of emitting gas coexist in one object,
the observed spectrum contains the contribution from all of them.
Therefore, although some models better fit high ionization lines and other
low ionization lines, the most probable model must consistently account
for all the lines and the continuum.

\subsection{The line spectrum in the UV}

In Table 1 the models are described and in Table 2 the calculated line
ratios  (to CIV 1533+1548 = 100) are compared  with the observed 
data from ZSH. The arrows indicate that the multiplet is summed up.

FWHM of line profiles from the observations are listed
in column 3 of Table 2. 
Penston et al (1983) measured  in 1978 a terminal velocity of the wind
 of about 75 \kms, indicating that 
the wind decelerated from initial velocities of thousands of \kms
~to less than 100 \kms. Large features were actually not observed,
however, wide feet of the high level lines are not excluded by 
Nussbaumer \& Dumm (1997) in later spectra.

We consider here the most significative lines. 
The spectrum at 1978 contains lines from relatively high levels (OV], NV, etc)
down to neutral lines. As already noticed by ZSH, the broader the line,
the higher the level of the ion. The line widths range between 70 \kms 
~and 30 \kms. The profiles are separated into two groups, one with 
FWHM $\sim $ 40 \kms ~and the other with FWHM  up to $\sim$ 80 \kms.
It is known that broad and narrow profiles of single lines characterize
most of the symbiotic spectra. In RR Tel the two main widths are very 
close to each other and difficult to recognize.

Model REV1(Table 1) with  a shock velocity of 80 \kms ~and a density of 
$10^6$ \cm3  represents the reverse shock. This model nicely fits high 
ionization  lines (from levels higher than III). NIV] 1386 is 
overestimated by a  factor of 4, while [Mg V] 1324 is  underestimated,
but the identification of this line is doubtful (ZSH). 
An alternative model for the reverse shock is REV2 characterized by a
lower density ($10^5$ \cm3), corresponding to a different location
of the emitting clump in the system.

Model EXP represents the shock propagating towards the red giant, and
is characterized by  a velocity of 45 \kms and a density of 2 $\times 10^6$ \cm3 as 
indicated by the observations. The absolute intensity of CIV is  extremely 
low for \Vs $<$ 50 \kms and U $<$ 0.01, therefore, the
line ratios to CIV =100 are very high for this model. However, these
are necessary conditions to obtain strong neutral lines. In fact, this  model 
is  constrained by the ratio of the O I/ C II line intensities.
The observed O I lines are relatively strong and show that the 
emitting clumps  should contain a large region of  cool 
recombined gas (see \S  6).

In the last two columns of Table 2  the composite models  SUM1 
and SUM2 are listed. They  result from the weighted sum of the contributions
of the expanding shock (EXP) spectrum and that emitted by  the reverse
shocks REV1 and REV2, respectively. The relative weights adopted 
are given in the bottom of Table 1.
The fit presented in Table 2 is quite good, taking into account 
the disomogeneity of the emitting matter in the colliding region and the
uncertainty of some atomic data in the code.
The spectrum is so rich in number of lines that the fitting
procedure  constrains the model, thus providing the most probable 
description of the system.
It seems that the best fit to the observed data is obtained by model SUM1.

The relative abundances  calculated phenomenologically  for
models SUM1 and SUM2 are in agreement  with  those adopted in the model of 
Hayes \& Nussbaumer (1986) within a factor of 2 (Table 1),
except for Mg/H which is higher by a factor  $>$20 .

\bigskip

Model XX, which is also listed in Table 1, is characterized by 
a high \Vs  ~and a high \n0 and will be discussed  in the context 
of the continuum radiation in the next section.
In fact, the full modeling of an object by the calculation of the
emitted spectra implies the cross-checking of the results obtained
for the line spectra with those obtained for the spectral energy 
distribution (SED) of the continuum. This procedure is necessary
because it  sometimes reveals that models other than those indicated
by the fit  of the line spectra must be taken into consideration.

\subsection{The continuum}

Model calculations are compared with the observations in Fig. 4.
The  observed data are taken from Nussbaumer \& Dumm (1997) in the 
wavelength range  between 1280 and 2600 \AA  ~and from Kenyon et al (1986)
in the IR. 
RR Tel has been detected as an X-ray source in 1978 by Einstein (Kwok \& Leahy
1984) and in 1992 by ROSAT (Jordan, Murset, \& Werner 1994).
The data in the X-ray are taken from  Fig. 3 of
Kwok \& Leahy (1984).
The SED of the continua calculated by models REV1 and EXP are represented
by solid lines and  dotted lines, respectively. For each model
besides the curve representing bremsstrahlung emission from the gas,
reradiation by dust is represented by the curve in the IR ($\nu < 10^{14}$ Hz).
Generally, the temperature of the grains downstream follows the trend of 
the temperature of the gas (Viegas \& Contini 1994, Fig. 3) therefore,
the location of the dust reradiation peak depends on the shock velocity. 
On the other hand, the intensity of the dust radiation flux relative to
bremsstrahlung emission depends on the dust-to-gas parameter, d/g.

The fit of model REV1 to the data in the IR yields to  d/g = 5 $\times 10^{-15}$,
whereas  for model EXP to d/g = $10^{-14}$. Since model EXP represents 
a shock front closer to the giant, a higher d/g implies a dustier 
atmosphere.

Fig. 4 shows that  the radiation flux  calculated by model REV1 strongly
prevails on that calculated by EXP, and that some absorption is present
in the low frequency domain.

As for the X-ray domain ($10^{16}$ - $10^{18}$ Hz), the data show  
emission from a  gas
at a temperature between $10^6$ and $10^7$ K .
Jordan et al. (1994) suggested that such a plasma could be produced by
a fast wind with velocity of $\sim$ 500 \kms.
This emission  cannot be  obtained either by 
the models which fit the emitting lines  or by a black body spectrum
corresponding to a colour temperature of 1.4 $\times 10^5$ K (dot-dashed line
in Fig. 4). Actually, bremsstrahlung  emission calculated by a model 
characterized by a high shock velocity (model XX in Table 1) nicely 
fits the data at higher frequencies (long dashed lines in Fig. 4). 
Notice that the  point at 
$\nu$ = 3 $\times 10^{16}$ Hz is below  the  calculated curve 
since it is affected by absorption of the interstellar medium (ISM)
(Zombeck 1990).

The high velocity   model  XX (\Vs = 500 \kms) could explain the 
wide foot underneath the nebular emission of CIV, NV, 
and HeII lines observed by  Nussbaumer \& Dumm (1997).
However,  since large FWHM are not observed in the line 
profiles, the contribution of the spectrum calculated by the 
high velocity model to the line fluxes should be negligible. 
The line  intensities calculated by model XX 
depend strongly on the geometrical thickness  of the emitting clump (d).
To obtain line intensities lower than those from model REV1, d  
should be lower than  6 $\times 10^{13}$ cm.

A high \Vs  ~is justified  in a  colliding scenario, as the wind collision
 yields to the disruption of the colliding matter
and to the formations of clumps. The fast wind will propagate
almost unaffected in the voids between the clumps.

An alternative scenario is represented by  model X0 
(short-dashed lines in Fig. 4) characterized
by \Vs = 500 \kms, but with a  lower density (about $10^5$ \cm3). 
The  downstream region at high temperature in the clumps
is rather extended ( $>$ 5 $\times 10^{14}$ cm) and  leads to the emission which 
fits the high frequency data  and to 
the requested almost negligible line fluxes.

\bigskip

\subsection{Variability on short time scale}

The photometric lightcurve  of RR Tel in 1978 is characterized by
large fluctuations in the visual band (see \S 2). Penston et al (1983) 
addressed this problem but
could not conclude whether these variations are caused by line or 
by continuum variations. Actually, [Cl III] 5539, [OI] 5577, [FeVI] 5631, 
and [NII] 5755 are the strongest lines which enter the 
broad band visual filter. 

In our model the shocks propagate throughout disrupted matter,
 heating and ionizing clumps of different geometrical width.
We present in Fig. 5 the calculated flux of these lines and 
of the continuum as a function of the distance from the shock
front for model REV1. This distance represents  the geometrical
width of the clumps since the models are matter bound. 
It can be seen that, particularly,  between $10^{13}$ and $10^{14}$ cm 
the line flux increases with distance from 
the shock front more than the continuum.  
This is a critical range for the thickness
of the clumps and within this range the line
variability dominates the variability of the continuum
in the light curve.

\section{The spectra in 1993}

We have calculated the spectra in 1993 in the frame of the 
scenario proposed in \S 2. The models which better
fit the observations are given in Table 3 and the line ratios are
compared with the observations in Tables 4a and 4b.
 
\subsection{The UV spectrum}

In the 1993 UV spectrum the high ionization lines dominate, and the 
OI and CII lines are still present, although weakened.
Therefore, a region of  gas emitting  low ionization-level and neutral
lines must still be  present. 
This corresponds to a model with low \Vs,
low or even absent U, and a large zone of emitting gas at low temperature.
Similar conditions were adopted to fit the 1978 spectrum (model EXP).
In other words, the contribution of the expanding shock to
the emitted spectrum is unavoidable. 
Notice that the flux intensity of CIV has decreased by a
factor of $ >$ 2 with respect to the 1978 flux.
In Table 4a  line intensities relative to CIV = 100 are listed  
together with the  FWHM from ZSH and the observed line ratios
(columns 3 and 4). Two models for the
reverse shock follow in columns 5 and 6, and three models
representing the expanding shock are given in columns 7, 8, and 9.
In the last three  columns of Table 4a the composite models 
sum1, sum2, and sum3  (which better fit the observations) are
presented. The relative weight of the  single contribution in these
models are listed in the bottom of Table 3.

The best  fit to the 1993 high level line spectrum, even if rough, 
is obtained by model rev1 which is characterized by a high \n0. 
In particular, the calculated intensity of [MgVI] 1806 line increased
by a factor of  1.9 from 1978. 
This trend agrees with the data listed by ZSH (Fig. 2).
The other
parameters of model rev1 (Table 3) suggest that, compared with 1978, 
the shock velocity has slightly decreased, the densities are higher, and 
U is lower than for model REV1 (see Table 1). 
A similar situation was found in the evolution 
of HM Sge (Formiggini et al 1995). 

On the other hand, comparing the FWHM of the low ionization and neutral 
lines between 1978 and 1993,  the expanding shock has maintained the same 
velocity, or  even increased slightly.
Three different  models are considered, because, depending on the 
location of the shock front, the preshock density can be very different.
We have adopted the same velocity for all the low level lines even if
the FWHM of these lines (Table 4b) show a wide range
of values from 17 \kms of [OI] 6300 to 83 \kms of [NII] 6584.

Model exp1 represents the case in which the shock front is still near the
giant star (Fig. 3b) and a high density characterizes the preshocked gas.
The expanding shock has maintained roughly the same velocity
compared to that in 1978 (model REV in Table 1).
In 15 years the shock could have overcome the red giant and be now propagating
in the giant atmosphere on the side opposite to the WD, where the
density gradient is decreasing. 
It is therefore predictable that radiation from the WD  should be
very low or even  absent because of dilution and/or screening effects.

Two other models seem acceptable. 
In model exp2~  the shock has reached a larger distance from the
giant on the side opposite to the WD (Fig. 3b). At this distance, 
the density is lower and the shock velocity is higher because the shock
accelerates while propagating through the decreasing density gradient in the
giant atmosphere.  The relative abundances are the same as those adopted 
for the other models because the shock is propagating within the system.

Model exp3 represents a different situation: the expanding 
shock colliding face-on-tail with the wind from the giant 
reached the outskirts of the system (Fig. 3b) and  is now propagating
in the ISM characterized by a lower density. The relative abundances
are close to the solar ones because mixing is rather strong 
at those large distances. 
In this model  the shock  could have traveled in about 15 years
to a distance  beyond 2.3 $\times 10^{15}$ cm.
The geometrical thickness of the emitting slab is large. 
The radiation flux from the hot star is not prevented
from reaching the inner edge of the slab but is low.

Notice that the three models exp1, exp2, and exp3 contribute only to
low level lines in the UV spectrum which are few compared to the numerous
high level lines. Therefore, we will constrain the models by the fit of
the optical spectrum where the low level lines are as numerous as the
high level ones. The consistent fit of line spectra in the different 
frequency ranges implies cross checking  of one
another until a fine tuning of all of them is found.

\subsection{The spectrum in the optical range}

In Table 4b observed line intensities in the optical range are compared
with model calculations. These lines are those which are  common to the
 observations of McKenna et al
(1997) and the SUMA code. The lines in the optical range are given 
relative to \Hb = 1. 
For consistency, the same models used to fit the line in the UV are adopted. 
However, model rev2 is omitted. 

Notice that [OII] 3727, even if low, is present in the spectra. 
Considering that [OII] has a critical density for collisional
deexcitation $\geq$  3 $\times 10^{3}$ \cm3,  the density in the
emitting gas cannot be much higher.
The electron density downstream decreases with recombination (\S 6), therefore,
even preshock densities higher than the critical one can fit the [OII]/ \Hb  ~line ratio.
On this basis, model exp3 is more acceptable than exp1 and exp2.

The calculated [OIII] 4959+5007 is  rather high compared with the observed 
line. A lower weight of model exp3 in the weighted sum (sum3) could improve
the fit of [NII] 5754 \Hb ~and [OIII]/ \Hb,  but definitively spoil the fit
 of [OII]/ \Hb.

Interestingly, a low R$_{[OIII]}$=[OIII] 5007 / [OIII] 4363 ratio coexists 
with  a relatively high [OII] 3727 / \Hb ~ratio. Low R$_{[OIII]}$ is generally
due to either high density or high temperature of the emitting gas.
High temperatures are  excluded because the shock velocity is low. Therefore, 
high densities are indicated by R$_{[OIII]}$ and low densities by [OII]/ \Hb.
This peculiarity of the spectrum strengthens the hypothesis of a
composite model. In fact, the low [OIII] 5007 / [OIII] 4363 ratio
comes from the rev1 model while the relatively high [OII] 3727 / \Hb ~ratio
~comes from model exp3.
The [OI] 6300 / \Hb ~line ratio, even if not very high, indicates that a 
large zone of neutral oxygen is present and is 
consistent with the large d (2 $\times 10^{15}$ cm) of model exp3.
The relatively high [OI]/ \Hb ~line ratio appears also in  model sum2
if a weight of 100 (see Table 3) is adopted for model exp2. This is rather 
high and  definitively spoils the fit of most of the other line ratios.
The composite model  sum1  nicely fits most of the optical-near IR lines,
however, [OI]/ \Hb ~and [NII] 6584+/ \Hb ~are largely overestimated.

In the composite model sum3, model exp3 has been given a weight three times that of model rev1.
This is acceptable
considering that  the emission from the nebula between the components is
certainly smaller than that from the nebula encircling the system.

Following the analysis of the optical spectrum, the composite model sum3 
is selected. This model also   fits  the UV spectrum best, 
even if the OI 1304 / CIV line ratio is underestimated (Table 4a).

\section{The  RR Tel system in the epochs  1978 and 1993}

After selecting the composite models which best fit the spectra observed
in each of the  epochs, a more quantitative description of the system can be given.

The results of model calculations give  a rough physical
picture of RR Tel in 1978.
The distance $\rm r_{rs}$ of the reverse shock from the WD can  be 
calculated from  :

U n c =  $\rm N_{ph}$ $\rm (R_{wd} / r_{rs})^2$   \,\,\,\,\,\,\,\,\,\ (1)

\noindent
where $\rm N_{ph}$ is the Plank function corresponding to \Ts = 1.4 $\times 10^5$ K,
and the WD radius $\rm R_{wd}$ = $ 10^9$ cm.
Adopting the parameters of model REV1, 
 $\rm N_{ph}$ = 3.4 $\times 10^{26}$ $\rm photons \, cm^{-2} \, s^{-1}$, and 
considering  the density n = 4 \n0 at the leading  edge of the clumps
due to the adiabatic jump, $\rm r_{rs}$ = 1.1 $\times 10^{14}$ cm.

Applying equation (1) to the expanding shock (model EXP), the distance between 
the wd  and the low velocity slab ($\rm r_{es}$) can be  calculated. 
In this case the density n is given after compression downstream (n = 3.2 $\times 10^7$ \cm3)
and $\rm r_{es}$ results 1.37 $\times 10^{15}$ cm.
The expanding shock is very close or even overpassing the red giant, considering the
wide interbinary separation ($\sim 10^{15}$ cm).

In 1993 the reverse shock is at about the same distance from the wd as in 1978.
On the other hand, from equation (1), the expanding shock represented by model exp3 
has reached a distance of about 5 $\rm 10^{15} $ cm corresponding to
the outskirts of the system (\S 5.1) 

As for the X-ray,  the detection by ROSAT in 1992 (Jordan et al 1994) 
indicates that
the high velocity wind (500 \kms) component is unchanged from 1978 (\S 4.2).

We compare now the physical conditions in the emitting gas
at the two epochs.

The profiles of  the fractional abundance of the most significant ions 
throughout the clumps 
in the two epochs are compared in Figs. 6a (1978) and 6b (1993).
In both figures the WD is on the left.
 The diagrams on the left
represent the clump downstream  of the reverse shock and the
diagrams on the right the clump  downstream of the expanding shock.
Notice that the black body radiation from the hot star reaches the very
shock front edge in the reverse shocked clumps and the edge opposite
to the shock front in the expanding ones.
To better understand the physical conditions in the emitting regions
the profiles of  the electron temperature, \Te , 
and of the electron density, \Ne, in a logarithmic scale  are also given
in the  top of the  figures.

First, we focus on models REV1 and rev1 (left side of Figs. 6a and 6b), 
the spectra of which largely prevail in the
  weighted sums of both epochs.
It can be noticed that $\rm He^{+2}$/He and $\rm C^{+4}$/C are very high
throughout  the clump geometrical width. The dotted vertical lines in Figs. 
6a and 6b
show the actual geometrical thickness of matter bound models. For comparison
the radiation bound
model results are also shown beyond these lines. 
The observational data show that CIV and HeII line intensities prevail
and that NV/CIV, HeII/CIV, and, - in a reduced
way - also OV/CIV line ratios, increase from 1978 to 1993 (see Tables 2 and 4).
In the same time CIV absolute flux decreases by a factor of $>$ 2.
 The fit to these observations
can be obtained  by slightly reducing the shock velocity and the ionization parameter,
and by increasing the preshock density, but, particularly,
the geometrical thickness of the clump must be reduced by a factor larger 
than 2 (Tables 1 and 3).

Regarding the  expanding shock, the physical 
conditions across the clumps in 1978 (model EXP) and 1993 (model exp3)
 are very different 
(Figures 6a and 6b diagrams on the right side of the figures).
The clump is divided in two halves by the vertical solid line.
The x-axis scale is  logarithmic and symmetric
to  have a comparable view of the two parts of the clump. The shock front is on
the right while the edge photoionized by the WD radiation is on the left.
In 1993 the electron density never exceeds $10^5$ \cm3.
The region corresponding to the $\rm O^{+0}$ ion prevails in 1978
throughout the whole clump, while in 1993 is reduced to the shock
dominated region.

\section{Conclusions}

In the previous sections we presented a model for RR Tel
at two epochs. The first one (1978) is characterized
by  large fluctuations in the light curve and in the
line intensities on very short time scales. 
 In the second epoch (1993) the system recovered
the fading trend. This epoch  is considered 
with the aim of investigating the evolution of the
system.

After the outburst, two shocks are present :
the reverse shock  propagates in the direction of the WD 
and the other one expands towards or beyond the giant.
A good fit of the observed emission line
spectra and continuum in 1978 is provided by a composite model,
where the reverse shock is characterized by a velocity of
$\sim$ 80 \kms ~and the expanding one by a velocity of 
$\sim$  45 \kms. 
A high velocity (500 \kms) wind component is 
revealed from the fit
of the SED of the continuum in the X-ray range in 1978,
but it is quite  unobservable in the line
profiles if the geometrical thickness of the clumps is lower than 
6 $\times 10^{13}$ cm.

The large fluctuations observed in the 1978 light curve result
from  line intensity  rather than from  continuum variation.
These  variations are explained by fragmentation of matter at 
the time of head-on collision of the winds from the two stars.

The results of our modeling show that in 1993 the reverse shock
 velocity has slightly decreased (70 - 50 \kms) and  the expanding
shock with velocity between 50 to 100 \kms has
overcome the symbiotic system and is propagating    
in the nearby ISM. The decrease in the absolute flux of  CIV is
justified by the reduced shock velocity and ionization parameter.
However, the geometrical thickness of the emitting clumps is
the critical parameter which can explain the short time scale
variabilities of the spectrum  and the trend of slow line
intensity decrease.

Relative abundances of carbon, nitrogen, oxygen, and silicon, to hydrogen, 
calculated by the present model, are in agreement
with those  assumed by the model of Hayes \& Nussbaumer (1986).

Finally, although our modeling is rather simplistic
it shows that shock models can contribute to a better understanding of the
outburst evolution of RR Tel. 

\bigskip

\noindent
Acknowledgements

We are grateful to an anonymous referee for 
helpful  comments and to G. Drukier for reading the manuscript.

\newpage

{\bf References}

\bigskip

\vsize=26 true cm
\hsize=14 true cm
\baselineskip=18 pt
%
\def\ref {\par \noindent \parshape=6 0cm 12.5cm 
0.5cm 12.5cm 0.5cm 12.5cm 0.5cm 12.5cm 0.5cm 12.5cm 0.5cm 12.5cm}

\ref Aller, L.H., Polidan, R.S., \& Rhodes, E.J.  1973, Ap\&SS, 20, 93
\ref Aufdenberg, J.P. 1993 ApJS 87, 337
\ref Bateson, F.M. 1995, private communication
\ref Bohigas, J., Echevarria, J., Diego, F., Sarmiento, J. A., 1989, MNRAS, 238, 
1395
\ref Contini, M. 1997, ApJ, 483, 886
\ref Feast, M.W. et al. 1983, MNRAS, 202, 951
\ref Formiggini, L., Contini, M., Leibowitz, E.M. 1995, MNRAS, 277, 1071
\ref Girard, T., Willson, L. A., 1987, A\&A, 183,247
\ref Hayes, M.A. \& Nussbaumer, H. 1986, A\&A, 161, 287
\ref Heck, A. \& Manfroid, J. 1985, A\&A, 142, 341
\ref Jordan,S., Murset, U., \& Werner, K. 1994, A\&A,  283, 475
\ref Kenyon, S.J., Fernendez-Castro, T., \& Stencel, R.E. 1986, AJ, 92, 1118
\ref Kenyon, S.J. et al 1993, AJ, 106, 1573
\ref Kwok, S.  \& Leahy, D.A.  1984, ApJ, 283, 675
\ref McKenna, F. C. et al. 1997, ApJS, 109, 225
\ref Murset, U., Nussbaumer,H.,Schmid, H.M., \& Vogel, M. 1991, A\&A, 248, 458
\ref Nussbaumer, H. \& Dumm, T.  1997, A\&A, 323, 387
\ref Nussbaumer, H., \& Schild, H. 1981, A\&A,  101, 118
\ref Nussbaumer, H., Walder, R., 1993, A\&A, 278, 209       
\ref Penston, M.V.  et al. 1983, MNRAS, 202, 833
\ref Thackeray, A.D. 1977, MNRAS,  83, 1
\ref Viegas, S. M., Contini, M., 1994, ApJ, 428, 113 
\ref Viotti, R.  1988 in "The Symbiotic Phoenomenon" eds. J.Mikolajewska et al.
Kluwer Academic Publishers, p. 269
\ref Wallerstein, G., Willson, L. A., Salzer, J., Brugel, E., 1984, A\&A, 133, 137
\ref Withelock, P.A. 1988 in "The Symbiotic Phoenomenon" eds. J.Mikolajewska et al.
Kluwer Academic Publishers, p. 47
\ref Zombeck, M.V.  1990 in "Handbook of Space Astronomy and Astrophysics"
Cambridge University Press, p. 199
\ref Zuccolo, R., Selvelli, P., \& Hack, M.  1997, A\&AS, 124, 425 (ZSH)

\newpage

{\bf Figure Captions}

\bigskip

Fig.1 :

The light curve of RR Tel between the years 1944 and 1993.

\bigskip

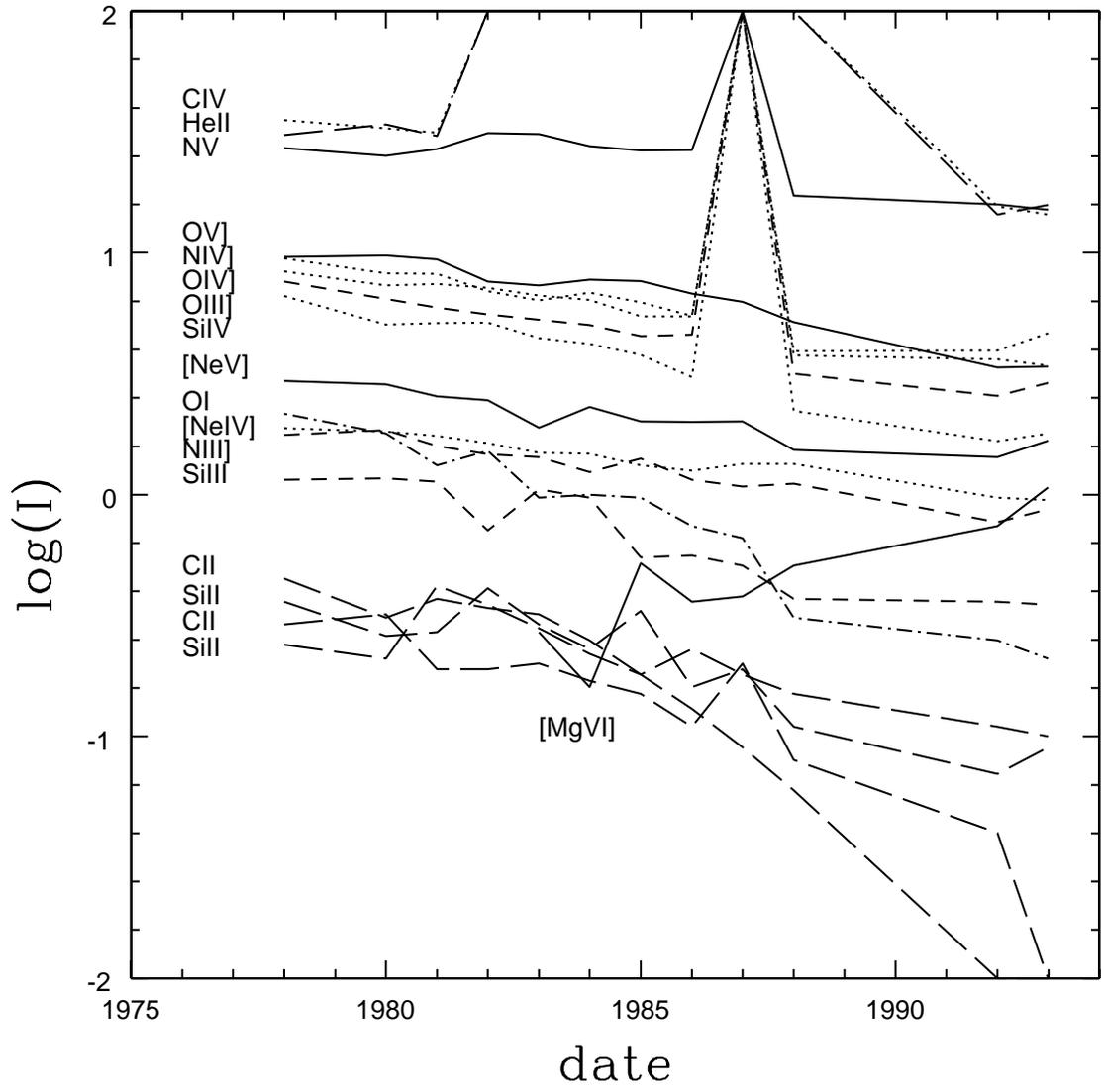
Fig. 2 :

The evolution of some significant line  intensities between 1978 and 1993.
from ZSH Table 1.
Saturated fluxes are arbitrarly set at log(I) = 2.

\bigskip

Figs. 3 :

Simple sketch of the model for RR Tel in 1978 and in 1993.
The asterisk indicates the WD and the full circle the cool
giant.  Shock fronts are indicated by solid lines. Dotted and 
long-dashed lines indicate the clump edge opposite to the shock front. 
Dashed arrows represents the radiation from the WD.
a) 1978; b) 1993

\bigskip

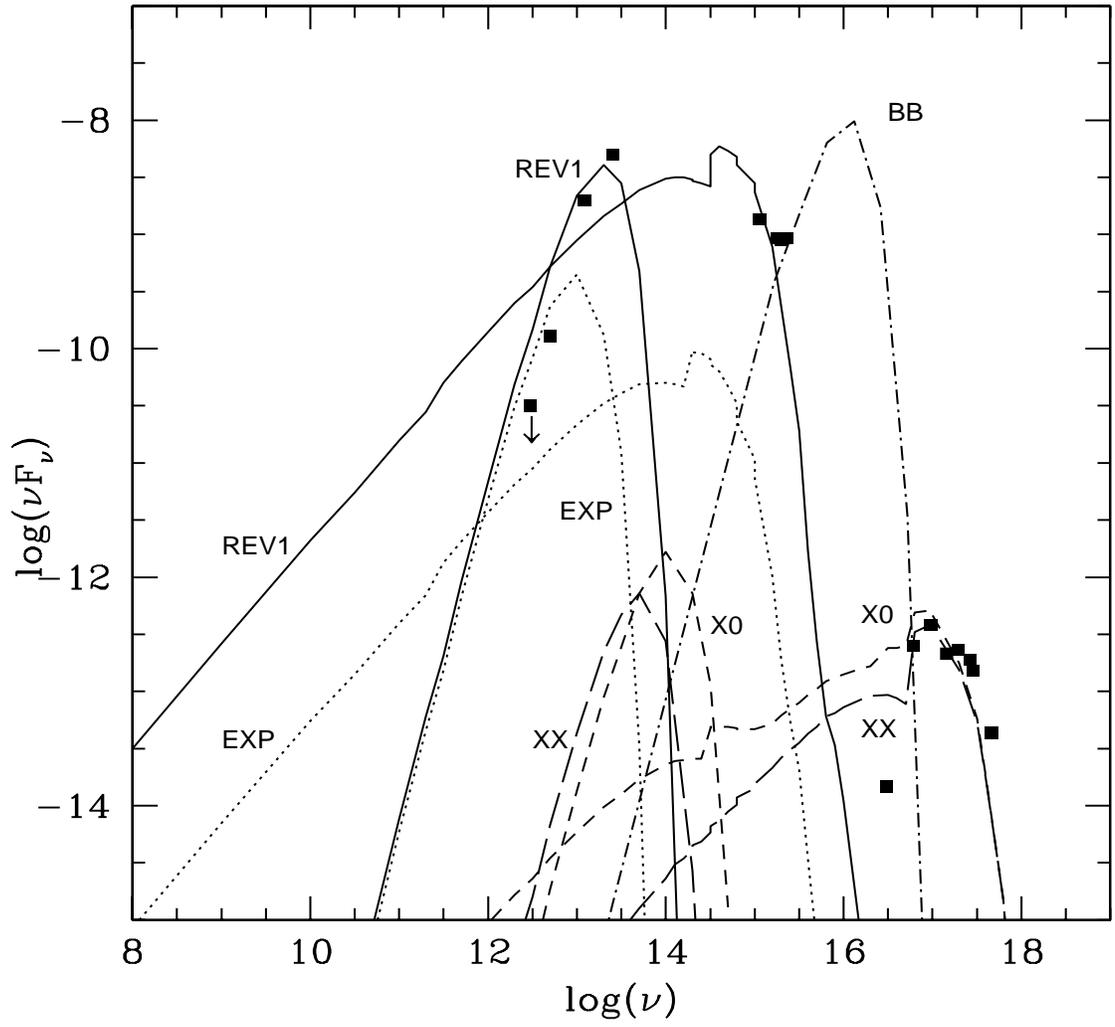
Fig. 4 :

The SED of the continuum in the year 1978.
Full squares represent the observations and the curves
represent model calculations (see text).

\bigskip

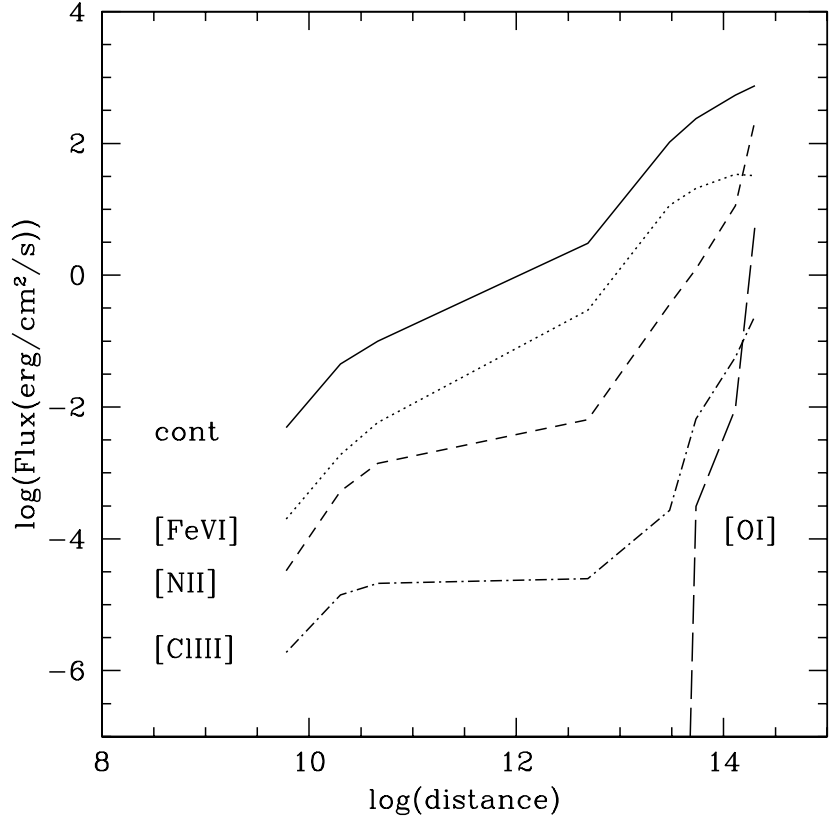
Fig. 5 :

The intensity  of some lines in the visual band  calculated as a function 
of the distance from the shock front.

\bigskip

Figs. 6 :

The distribution of the ions corresponding to the strongest line
throughout the clumps  (see text).
a) 1978, b) 1993.

\newpage

\topmargin 0.01cm
\oddsidemargin 0.01cm
\evensidemargin 0.01cm

\begin{table}
\centerline{Table 1}
\centerline{The parameters of Table 2}
\begin{tabular}{lll llll l l l l }\\ \hline   \\
     &REV1&REV2 &EXP& XX&HN(1986) \\                
\   \Vs (\kms)   &80.& 85.&45. &500&-               \\
\   \n0  (\cm3)  &1(6)& 1(5) &2(6)& 1(6)&-              \\
\   \B0  (gauss)  &1(-3)&1(-3)&1(-3)  &1(-3)&-                 \\
\   \Ts  (K)  &1.4(5)&1.4(5)&1.4(5) &1.4(5)& 2.(5)                  \\
\    U        & 0.2&0.2 &2.(-4)&0.2&-       \\
\ d (cm) & 5.5(13)&8.15(14) & 1.(14)& $<$ 5.9(13)&-                    \\
\ d/g & 1(-14)&5(-15)&1(-14)&  5(-15)&-  \\    
\  CIV (\erg) & 9.4(3)&1.(3) &5.9(-4) & $<$0.016  &-     \\
\ He/H& 0.1 &-&0.1 &-& 0.2 \\
\ C/H &  3.3(-4) &-&3.3(-4) &-& 3.5(-4)             \\
\ N/H &  5.4(-4) &-&5.4(-4)  &- & 3.(-4)           \\
\ O/H &  8.6(-4) &-&8.6(-4)   &-& 9.(-4)           \\
\ Mg/H & 6.9(-4) &-& 6.9(-4) &-& 3.(-5) \\
\  Si/H &  5.3(-5) &-& 5.3(-5)&-& 3(-5)              \\
\ W (SUM1) & 1. & - & 1. & - & - \\
\ W (SUM2) & - & 1. & 1. & - & - \\
 &&&&&& \\ \hline \\
\end{tabular}
\end{table}

\begin{table}
\centerline{Table 2}
\centerline{line ratios to CIV = 100 (1978)} 
\begin{tabular}{lll llll l l l l }\\ \hline 
   &line & FWHM & obs. &REV1& REV2 & EXP& SUM1 & SUM2 \\       
   &     & \kms &ZSH & &  & & & \\ 
\ 1199 & [SV]& 47.  & 1.51? &0.45&0.4 &-& 0.45& 0.4           \\
\ 1206 & SiIII & 55.& 1.15 &1.2&0.85 & 4.9(5)&1.2  & 1.14   \\
\ 1218 & OV] & 59. & 9.63&9.6  &9.6 & - &9.6& 9.6     \\
\ 1236 & NV &61 &27.14&31. &32.&- &31. &32.      \\
\ 1242 & NV & 51. &13.67 &-&-&-&- &- \\
\ 1264 & SiII & 45. & 0.08&0.016 &0.01 & 3.1(6)&0.2  & 1.8 \\
\ 1302 & OI & 51.&1.41&- &-  &3.7(6) &0.23& 2.2 \\
\ 1305 & OI & 39. &0.30 &$\uparrow$&$\uparrow$ &$\uparrow$&$\uparrow$&$\uparrow$ \\
\ 1306 & OI &48. &2.16 &$\uparrow$&$\uparrow$ &$\uparrow$&$\uparrow$&$\uparrow$ \\
\ 1324 & [MgV] &72.& 0.33&0.3&0.3& -&0.3  &0.3      \\
\ 1334 & CII & 40. & 0.29 &0.07& 0.12&1.8(5)&0.1 & 0.23  \\
\ 1335 & CII & 65. &0.45 &0.03&0.02&7(6) &0.45 & 4.1   \\
\ 1393 & SiIV & 49. & 6.64b&11.&10. & -&11. &10.      \\
\ 1398 & SIV] & 43 & -&0.3&0.28&-&0.3&0.28  \\
\ 1401 & OIV] & 47.  & 8.37b&7.8 &6.9 & 1.1(4) &7.8& 6.9 \\
\ 1404 & OIV]+ SIV] & 47. & 4.73b&4.6 &4.0&- &4.7 &4.0  \\
\ 1407 & OIV] & 53.   & 1.34&1.2&1.2&- &1.2 & 1.2  \\
\ 1483 & NIV] & 49.  & 0.41&0.56&0.5&-  &0.56&0.5   \\
\ 1486 & NIV] & 42.  & 9.49b&43&37.& -&43.&37.  \\
\ 1533 & SiII & 66.  & 0.24&0.012& .012&7(5)&0.06 & 0.42 \\
\ 1548 & CIV & 52.   & 66.55a&100 &100&100&100 &100 \\
\ 1551 & CIV & 50.  & 34.40a &$\uparrow$&$\uparrow$ &$\uparrow$&$\uparrow$&$\uparrow$ \\
\end{tabular}
\end{table}
\newpage
\begin{table}
\begin{tabular}{lll llll l l l l }\\ \hline
   &line & FWHM & obs. &REV1& REV2 & EXP& SUM1 & SUM2 \\       
   &     & \kms &ZSH & &  & & & \\ 
\ 1574 & [NeV] & 61.  &2.95 &0.17&0.11 & - &0.17& 0.11\\
\ 1601 & [NeIV] & 64.   & 1.88b&2.3&2.3&- &2.3 &2.3 \\
\ 1640 & HeII & 71. &30.70 & 32&36.& 1.6(7)&32. & 45.  \\
\ 1660 & OIII] & 38.   &2.57b&5.1&6.8 & -&5.1 & 6.8   \\
\ 1666 & OIII] & 36.   & 7.61 &$\uparrow$&$\uparrow$ &$\uparrow$&$\uparrow$&$\uparrow$ \\
\ 1718 & NIV & 78.  & 0.27 &0.25 &- &-&0.25&-&-\\
\ 1746 & NIII] & 48.  & 0.12 &\dop&$\downarrow$&$\downarrow$ &$\downarrow$& $\downarrow$\\
\ 1748 & NIII] & 45.   & 0.43 &\dop&$\downarrow$&$\downarrow$ &$\downarrow$& $\downarrow$\\
\ 1749 & NIII] & 41.  & 1.77b &4.2&8.&1.2(4)&4.2& 8. \\
\ 1752 & NIII] & 48.   & 0.72b &$\uparrow$&$\uparrow$ &$\uparrow$&$\uparrow$&$\uparrow$ \\
\ 1754 & NIII] & 44.   &0.37b &$\uparrow$&$\uparrow$ &$\uparrow$&$\uparrow$&$\uparrow$ \\
\ 1806 & [MgVI]& -     & -    &0.3 & 0.9 & - & 0.3 & 0.9 \\
\ 1808 & SiII & 45.   & 0.36 &0.05 &0.07 & 2.1(6)&0.18 & 1.3 \\
\ 1814 & [NeIII]& 43.  & 0.19 &0.07&0.07&1.8(4) &0.08& 0.07 \\
\ 1882&SiIII&-& 0.04  &$\downarrow$&$\downarrow$&$\downarrow$ &$\downarrow$& $\downarrow$\\
\ 1892 & SiIII & 40. &5.10a &6.6&5.5&1.5(6)&6.7 & 6.4 \\
 &&&&&&&&& \\ \hline \\
\end{tabular}

\end{table}

\topmargin 0.01cm
\oddsidemargin 0.01cm
\evensidemargin 0.01cm

\begin{table}
\centerline{Table 3}
\centerline{The parameters of Tables 4a and 4b}
\begin{tabular}{lll llll l l l l }\\ \hline 
     &rev1&rev2&exp1&exp2  & exp3   \\                
\   \Vs (\kms)  &60. & 70.&50. & 100.  &50.               \\
\   \n0  (\cm3)  &2(6)& 2(5) &5(6)& 3(5)&2(4)                  \\
\   \B0  (gauss)  &1(-3)&1(-3)&1(-3) & 1(-3)&1(-3)            \\
\   \Ts  (K)  &1.4(5)&1.4(5)&1.4(5) &  1.4(5)&1.4(5)                \\
\ U        & 0.1&0.1 & 0. & 0. &5(-3) \\
\ d (cm) & 2.9(13)&6.(12)   & $>$ 8.8(13) & 1.7(13)&2(15)         \\
\  d/g & 1(-14)&1(-14)&1(-14)& 1(-14)&1(-14)                          \\
\  CIV (\erg) & 4.4(3)&440.  &3.(-4)  &  12. &0.09               \\
\ \Hb (\erg) & 204. & & 139. &    0.12 &0.84 \\
\ C/H &  3.3(-4) &&3.3(-4) &3.3(-4) &3.3(-4)               \\
\ N/H &  5.4(-4) &&5.4(-4)  &5.4(-4) &1.4(-4)              \\
\ O/H &  8.6(-4) &&8.6(-4)   &8.6(-4) &6.6(-4)             \\
\ Mg/H&  6.9(-4) &&6.9(-4)   &6.9(-4) &6.9(-4)  \\
\  Si/H &  5.3(-5) && 5.3(-5) &5.3(-5) &5.3(-5)                \\
\ W  (sum1) & 1. & - & 1. & - & - \\
\ W  (sum2) & 1. & - & - & 100.& - \\
\ W  (sum3) & 1. & - & - & - & 3. \\

 &&&&&& \\ \hline \\
\end{tabular}
\end{table}

\begin{table}
\centerline{Table 4a}
\centerline{UV line ratios to CIV = 100 (1993)} 
\begin{tabular}{lll llll l l l ll }\\ \hline \\
&line & FWHM & obs. &rev1 &rev2 &exp1 &exp2 &exp3 &sum1 &sum2 &sum3   \\
&     & \kms &ZSH    & & & & && &  &  \\ 
\ 1199 & [SV]&  -   & 1.72  &  0.5 & 0.5&-&0.7&-&0.5&0.56 & 0.5    \\
\ 1206 & SiIII & -  & 1.14 &1.4 & 2. & 9.(6)&49.&- & 2.&11.6 & 1.4                \\
\ 1218 & OV] & 54. & 11.0&13. & 12.&- &1.2&-&13.& 10.5 & 13.                       \\
\ 1236 & NV &58 &49.2 & 41. & 36. &-&10.& 41.&- &34.3 & 41.                 \\
\ 1242 & NV & 52a &23.5  &-&- &-&-&-&-&- &- \\
\ 1264 & SiII &  -  & -   & 0.02 & 0.05 & 1.2(6) &1.2&23.7& 0.10  &0.27 & 0.02    \\
\ 1302 & OI & -  &0.39& 0.& 0.& 1.1(6) &0.11&2.36& 0.08  &0.02 & 1.4(-4)         \\
\ 1305 & OI &  -  &   - &$\uparrow$&\upa&\upa&\upa&\upa& \upa & \upa & \upa \\
\ 1306 & OI &50. &0.68  &$\uparrow$&$\uparrow$&\upa&\upa&\upa& \upa & \upa & \upa \\
\ 1324 & [MgV] &52.& 1.1 & 0.9 & 0105 & -&- & -&0.9&0.69 & 0.9               \\
\ 1334 & CII & -   & -    & 0.04 & 0.08 & 4.(4)&0.024 & 0.04& 298. &0.05 & 0.08               \\
\ 1335 & CII & 47. &0.32 &  -    & -    & -     &-    & -      & -   & - &-  \\
\ 1393 & SiIV & 52. & 5.8  &  13. & 14. & 9.6(5)&27.7 & 14.9&13.& 16.15 & 13.         \\
\  1398& SIV] & -    & -    & 0.24& 0.47& 2.2(3)&1.6& 0.24 &-&0.53 & 0.24                     \\
\ 1399 & OIV] &  49. &2.28b &-&-&-&-&- &-&-&- \\
\ 1401 & OIV] & 49.  & 15.2 & 10. & 9.3&-& 7. &- & 10.  & 9.36 & 10.  \\
\ 1402 & SiIV &  51. & 3.2& \upa & \upa &\upa& \upa & \upa  & \upa & \upa & \upa \\
\ 1404 & OIV]+ SIV] & 43. & 7.33 & 6.0 & 5.4 & - &4.&-& 6.& 5.6 & 6.     \\
\ 1407 & OIV] & 47.   & 1.98&1.7 & 1.5 & - &1.14&-& 1.7  &1.6 & 1.7       \\
\ 1483 & NIV] & - &0.52&$\downarrow$&$\downarrow $&\dop &\dop&\dop & \dop & \dop & \dop \\
\ 1486 & NIV] & 36.  & 11.1 & 54. & 47.& - &28.5& 54.&-& 48.5& 54.         \\
\ 1533 & SiII &  -   &  -  & 0.014 & 0.03 & 2.4(6) &0.46&228. & 0.18 &0.11  & 0.03      \\
\ 1548 & CIV & 50.   & 100   &100 &100&100&100 &100 &100 &100 & 100 \\
\ 1551 & CIV & 50.&$\uparrow$ &\upa &\upa &\upa& \upa & \upa & \upa & \upa & \upa \\
\end{tabular}
\end{table}
\newpage
\begin{table}
\begin{tabular}{lll llll l l l ll }\\ \hline \\
&line & FWHM & obs. &rev1 &rev2 &exp1 &exp2 &exp3 &sum1 &sum2 &sum3   \\
&     & \kms &ZSH    & & & & && &  &  \\ 

\ 1574 & [NeV] & 55.  &5.4  & 0.3 & 0.19 & -&2(-3) &-& 0.3& 0.23 & 0.3              \\
\ 1601 & [NeIV] & 56.   & 3.09 &3.3 & 2.7 & - &0.9& -&3.3& 2.8 & 3.3           \\
\ 1640 & HeII & 66. &51.3  & 40. & 34. &10.&0.24&-& 40. & 31.5 & 40.           \\
\ 1660 & OIII] & 41.   &3.26 & 3.3 & 4.8 & 1.3(4)&21.2 & 167.&3.3  & 7.13 & 3.3      \\
\ 1666 & OIII] & 38.a  & 9.44a&- &-&-&-&-&- &-&- \\
\ 1746&NIII] & 81.&0.13&$\downarrow$&$\downarrow $&\dop &\dop &\dop &\dop & \dop & \dop \\
\ 1748 &NIII]& 39.&0.49&$\downarrow$&\dop &\dop &\dop &\dop & \dop & \dop & \dop \\
\ 1749 & NIII] & 41.  & 2.8   &  3.& 5. & 4.7(5)&22.7 &32.4& 3.&7.22 & 3.        \\
\ 1752 & NIII] & 46. & 1.1  &$\uparrow$  & $\uparrow$ &\upa& \upa & \upa & \upa &\upa &\upa \\
\ 1754 & NIII] & 34. &0.42& $\uparrow$  & \upa &\upa& \upa & \upa & \upa & \upa & \upa \\
\ 1806 & [MgVI]& 0.64&3.47 & 1.2 & 1.5 & - & - & - & 1.2 & 0.9 & 1.2 \\
\ 1808 & SiII & 65.   & 0.29 & 0.06 & 0.13 & 2.9(7)&1.45 &210.& 2. & 0.36 & 0.07      \\
\ 1814 & [NeIII]& 33.  & 0.26 &0.05 & 0.07 & - &0.2& -&0.05  & 0.08 & 0.05                 \\
\ 1882&SiIII&-&  -   &$\downarrow$&\dop &\dop  &\dop  &\dop & \dop & \dop & \dop \\
\ 1892 & SiIII & -   &-     & 14. & 1.5 (7) &44.7& 15. &877. &14.&  14.2 & 14.              \\
 &&&&&&&&& &&\\ \hline \\
\end{tabular}

\end{table}

\begin{table}
\centerline{Table 4b}
\centerline{optical line ratios to \Hb = 1. (1993)} 
\begin{tabular}{lll llll l l l l }\\ \hline \\
& line & FWHM & obs &rev1&exp1&exp2& exp3 &sum1  & sum2 & sum3      \\ 
\ 3710 & SIII & 50. & 3.4(-3) & 9.(-4)& 7.(-4) &0.26& 1.2(-3)& 8.2(-4)&0.015&0.001 \\
\ 3729b+ & [OII] & 48. & 0.02 & - & 2(-3) & 0.22& 1.2 &8.2(-4)& 0.012 & 0.014 \\
\ 3759 & FeVII & 69. & 0.91 & 0.84 & - & 2.7(-3)&  -   & 0.5 & 0.79 & 0.84 \\
\ 3869b+ & [NeIII]& 43. & 0.79& 0.4 & - &3.1&2.5&0.23 & 0.55 & 0.42 \\
\ 4068+ & [SII] & 46. & 0.02 & - & 0.33 &0.6&1.7& 0.13& 0.03 &0.02   \\
\ 4114 & [FeII] & 32. & 4.9(-3) & - & 0.19 &0.16&0.36& 0.08& 8.8(-3)&4.3(-3) \\
\ 4363 & [OIII] & 40. &0.5 & 0.54 & - &5.&0.16& 0.32 & 0.85& 0.54 \\
\ 4625 & [ArV] & 45. & 5.(-3) & 0.019 & - &3.(-3)&-& 0.01 & 0.018& 0.019 \\
\ 4686 & HeII & 57. &    0.9 & 1.0 & - &0.02&0.37& 0.6 & 0.94 & 0.99 \\
\ 4740 & [ArIV] &67. & 0.013 & 7.2(-3) &-&6.(-3)&-& 4.2(-3)&7.1(-3)&7.(-3)   \\
\ 4893 & [FeVII & 65. & 0.05 & 0.12&-&-&-& 0.07&0.11& 0.12 \\
\ 4959+ & [OIII] & 59. & 0.35 & 0.6 & - &11.&17.2 & 0.35&0.29 & 0.53\\
\ 5676 & [FeVI] & 60. & 1.4(-3) & 0.06 & - &0.09& -& 0.03 & 0.06 & 0.06\\
\ 5754 & [NII] & 41. & 1.6(-3) & 2.3(-3) & 0.12 &5.5&-& 1.4(-3)&0.3 & 0.06 \\
\ 5876 & HeI & 41. & 9.(-3) & 7.4(-3) & 7.(-3) & 0.3&0.13 &7.2(-3)&0.02 & 9.(-3) \\
\ 6300+ & [OI] & 17. & 9.(-3)& - & 1.65 & 0.13&0.67 & 0.7& 7.2(-3)& 8.(-3) \\
\ 6310+ & [SIII] & 40. & 1.3(-3) & 1.5(-3) & 1.1(-3) & 0.42&1.8(-3) &1.3(-3)&0.024&1.5(-3) \\
\ 6584+ & [NII] & 83. & 5.(-3) & 4.(-4) & 0.54 &1.44&3.8 & 0.2 &0.08 & 0.01 \\
 &&&&&&&&&& \\ \hline \\
\end{tabular}
\end{table}

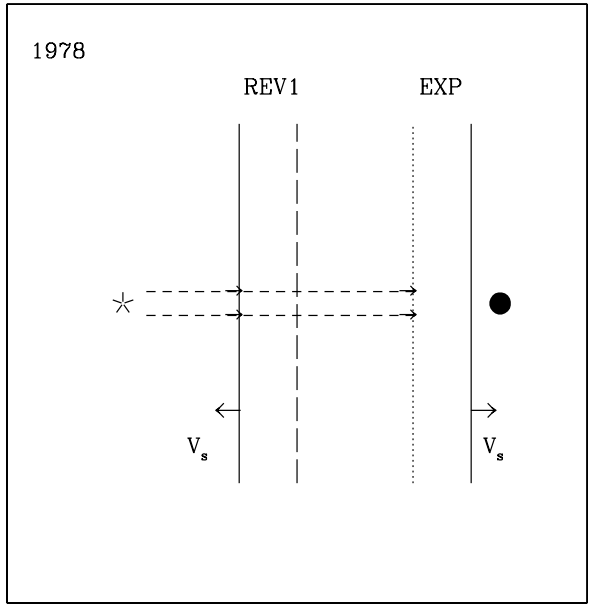
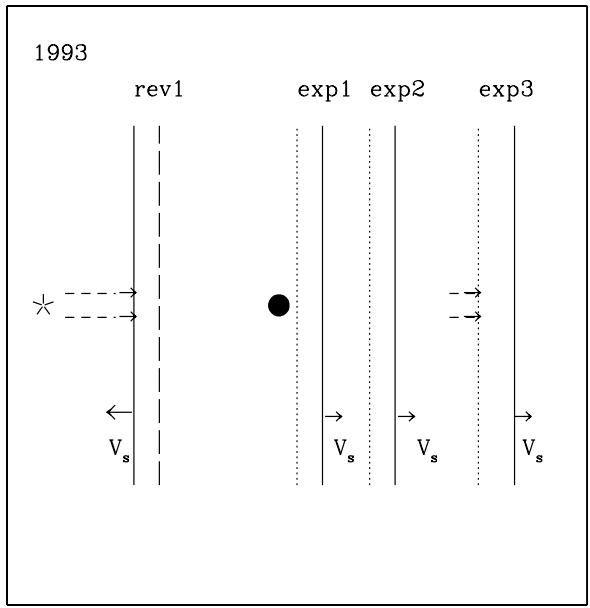
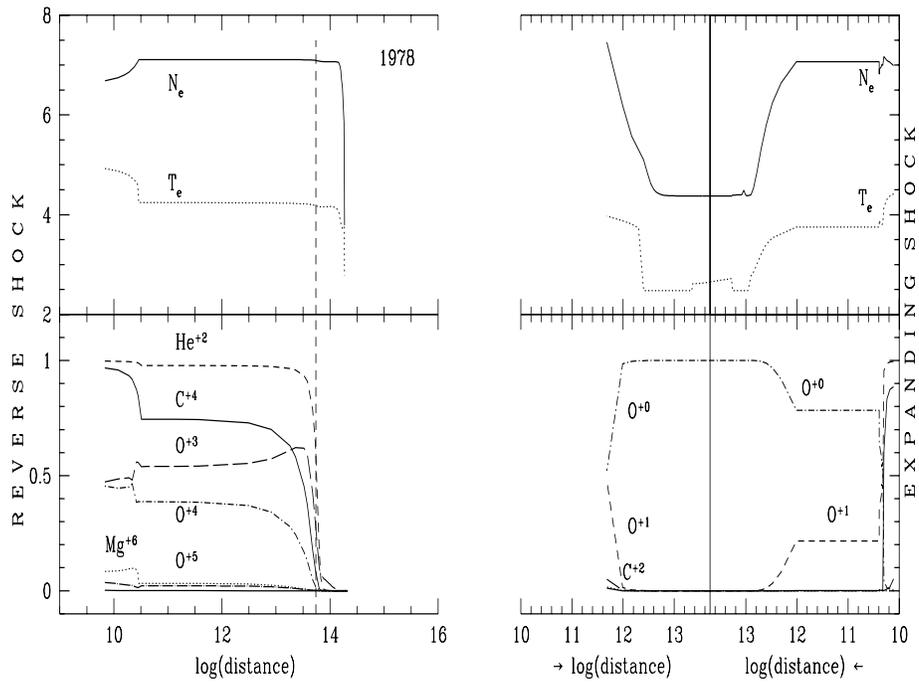
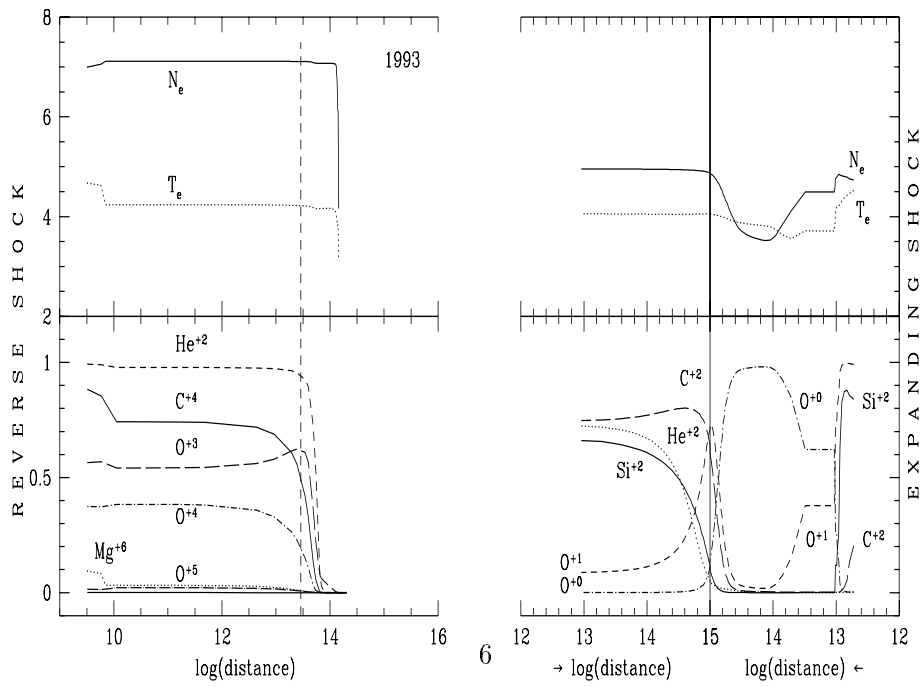
\end{document}